\documentstyle[12pt,aasms4]{article}
\tighten

\newcommand\be{\begin{equation}}
\newcommand\ee{\end{equation}}
\newcommand\jcd{Christensen-Dalsgaard}

\input epsf
\lefthead{Antia and Basu}
\righthead{High wavenumber solar oscillations}
\begin{document}

\title{High frequency and high wavenumber solar oscillations}

\author{H. M. Antia} 
\affil{Tata Institute of Fundamental Research, 
Homi Bhabha Road, Mumbai 400005, India}
\and
\author{Sarbani Basu}
\affil{Institute for Advanced Study, Olden Lane, Princeton NJ 08540,
U. S. A.}

\baselineskip 12 pt

\begin{abstract}
We determine the frequencies of solar oscillations covering a wide range
of degree ($100<\ell<4000$) and frequency ($1.5 <\nu<10$ mHz) using
the ring diagram technique applied to power spectra obtained from
MDI (Michelson Doppler Imager)  data. 
The f-mode ridge  extends
up to $\ell\approx3000$, where the line width becomes very large, implying
a damping time which is comparable to the time period. 
The frequencies of high degree f-modes are significantly different from
those given by the simple dispersion relation $\omega^2=gk$.
The f-mode peaks in power spectra are distinctly asymmetric and
use of asymmetric profile increases the fitted frequency bringing them
closer to the frequencies computed for a solar model.
\end{abstract}


\section{Introduction}

It is well known that solar acoustic modes with frequencies less
than the acoustic cut-off frequency of the photosphere
($\approx 5.3$ mHz) are trapped
below the solar photosphere. These modes have been successfully used to
probe the internal structure and dynamics of the Sun.
At higher frequencies the modes are expected
to propagate into the chromosphere, where they are likely to be
damped due to radiative dissipation. Nevertheless, a number of
observations (cf., Duvall et al.~1991; Fernandes et al.~1992; Kneer \&
von Uexk\"ull 1993)
have conclusively established that the p-mode ridges continue well beyond
the estimated acoustic cut-off frequency in the solar photosphere.
Various mechanisms have been put forward to explain the observed
high frequency p-modes.
Balmforth \& Gough (1990) suggested that the high frequency
modes may arise due to the reflection of waves at the transition layer
between the chromosphere and corona, where the temperature  and
hence the sound speed rises rapidly. In that case we would also expect
to observe the chromospheric modes  trapped in the
chromospheric cavity (Ulrich \& Rhodes 1977; Ando \& Osaki 1977)
expected to be formed around the temperature
minimum. Despite several attempts, there has not  been any
unambiguous detection of the chromospheric modes
(Duvall et al.~1991; Fernandes et al.~1992; Kneer \& von
Uexk\"ull 1993), and hence
Kumar \& Lu (1991) suggested that the high frequency modes arise
from interference between two traveling acoustic waves. 
It has been argued that
because of inhomogeneities in the structure of the chromosphere corona
transition layer the chromospheric cavity may not be sufficiently well
defined to support resonant modes (Kumar 1993).

Apart from the p-modes, f-modes have also been detected at high
degrees. These high degree f-modes are expected to be localized in
the chromosphere and hence can provide a
diagnostic for velocity fields prevailing there.
The frequencies of these modes are expected to be essentially independent
of the  stratification in solar interior and hence the difference
between the observed frequencies 
(cf., Libbrecht, Woodard, \&
Kaufman 1990; Bachmann et al.~1995; Duvall, Kosovichev \& Murawski 1998)
and those of solar models have been interpreted to be due to velocity
field or magnetic fields and other
effects (cf., Campbell \& Roberts 1989;
Murawski \& Roberts 1993; Rosenthal \& Gough 1994).
Although, Fernandes et al.~(1992) have measured
the frequencies of f-modes up to $\ell\approx3700$, 
the accuracy of these frequencies is limited
due to the short duration of their observations (2.4 hrs). Similarly,
using 8 hrs of observations, Duvall et al.~(1998) have calculated the
frequencies of f-modes up to $\ell\approx1800$ and they have interpreted
the decrease in observed frequency as compared to model frequency as
arising due to turbulent velocity fields. 
It would thus be interesting to
check the frequencies of these modes at higher degree as these are
expected to be localized in the chromosphere. 

The frequencies of these modes have been calculated by fitting a
symmetric Lorentzian profile to the power spectra. 
It has been
demonstrated that in general the peaks in power spectra are not
symmetric (Duvall et al.~1993; Toutain 1993; Nigam \& Kosovichev 1998;
Toutain et al.~1998) and the use of symmetric profiles may cause
the fitted frequency to be shifted away from the true value.
Thus, it would be interesting to estimate the frequencies using fits
to asymmetric profiles.

In this work we use the data obtained by 
Michelson Doppler Imager (MDI) on board the Solar and Heliospheric
Observatory  (SOHO) to measure frequencies of f and p-modes
at high degrees.  We have examined
power spectra that extend up to  degree    $\ell\approx 5000$ and
frequency $\nu\approx16$ mHz, to identify oscillatory modes with
degrees up to $\ell=4000$ and frequencies up to $\nu=10$ mHz.
Beyond these limits there is too little
power to fit the modes reliably. The rest of the paper is organized as
follows: we  describe the observations and analysis technique in
Section 2, while the results are described in Section 3.
Finally, we  summarize the main conclusions from our study in Section 4.

\section{The technique}

We adopt the ring diagram technique (Hill 1988; Patr\'on et al.~1995)
to determine the mean frequency of
solar oscillations at high degree ($\ell>100$)
(cf., Fernandes et al.~1992).
We use  MDI data and the MDI data-analysis pipe-line
to obtain three dimensional (3d) power spectra of solar oscillations
as described  by Basu, Antia \& Tripathy (1999). However unlike
Basu et al.~(1999) we have
not applied any temporal detrending to the time series since that
produces a broad peak at the frequency corresponding to that of the
detrending interval.

We have used both  full-disk and  high-resolution
Dopplergrams produced by MDI.
Full-disk Dopplergrams give the power spectra for modes
with degree up to $1500$, while
high-resolution Dopplergrams can produce spectra for modes with degrees 
up to $5000$.
These observations are taken at a cadence of 1 minute
thus giving a Nyquist limit of 8.33 mHz.
From the available full-disk Dopplergrams we have selected
24 intervals of 1536 minutes each 
covering a period from May 25 to June 21, 1996.
These correspond to longitudes centered at 90, 76, 60, 45, 30 and 15 
degrees for Carrington rotation 1909, and longitudes 360,
345, 330, 315, 300, 285, 270, 255, 240, 225, 210, 195,
181, 160, 150, 135, 120 and 105 degrees for Carrington 
rotation 1910. 
The slightly uneven distribution in the
longitudes is due to our attempt to avoid data gaps.
For each time series we have selected three regions
centered at the central meridian and at latitudes of $0,\pm15^{\circ}$,
covering
$256\times256$ pixels in longitude and latitude.
We have summed all  72 spectra to improve statistics.
These summed spectra effectively cover a time interval of 26 days, which
improves the statistics.

For the spectra obtained from high-resolution Dopplergrams
we have taken the sum over 24 spectra covering the available data
for March 1997 and from June to July 1997. 
The longitude and latitude distribution of the data used
was determined by the availability of observed images.
We used time series of $256\times256$ pixel images
centered at a latitude of 11$^\circ$ N and
longitudes of 200 and 185 degrees for Carrington rotation 1920,
at latitudes of 11$^\circ$ N and 18$^\circ$ N  for
longitudes 50, 35, 20 and 5 degrees for Carrington 
rotation 1923, and
for  longitudes 135, 120, 105, 90, 75, 60 and 45 degrees
for Carrington rotation 1924.
Although we tried to select periods with least data-gaps,
nevertheless the high-resolution data sets  have many more gaps than the
full-disk set and this affects the quality of the resulting spectra.
Since the high-resolution Dopplergrams do not cover the entire solar disk
there could be some error in estimating the spatial scale, which has
been obtained from the relevant parameters in the image headers.
The difference in frequencies obtained from the two spectra would give
an estimate of such systematic errors.
We have also used full-disk
Dopplergrams taken at higher cadence of 30 s to study higher frequency
modes that  extend to 16 mHz. For this purpose we have summed over
24 spectra, each obtained from time series of 1024 minutes,
covering the available data for June 1997. 

These summed  3d spectra are averaged in the azimuthal direction to
obtain 2d (two dimensional)
spectra in $k$ and $\nu$, where $k$ is the horizontal wavenumber.
For the full-disk spectra the Nyquist limit on individual spatial
components $k_x$ and $k_y$ corresponds to $\ell\approx1500$.
However, while taking the azimuthal averages we can extend the spectra to
about $\ell=2100$ (the maximum value of $k=\sqrt{k_x^2+k_y^2}$)
by averaging over partial and incomplete rings too
at fixed temporal frequency. This can introduce some systematic errors
for $\ell>1500$ because of spatial aliasing and averaging over only a
part of ring. However, comparison with results obtained using the
high-resolution Dopplergrams shows that this effect is not very significant.

We fit the sections at constant $k$ (or $\ell$) in the averaged spectra
to a Lorentzian profile of the form,
\be
P(\ell,\nu)={\exp{A_0}\over(\nu-\nu_0)^2+w_0^2}+
e^{b_1}[1+b_2(1-\nu/\nu_c)],
\ee
where the 5 parameters $A_0,\nu_0,w_0,b_1$ and $b_2$ are determined by
fitting the spectra using a maximum likelihood approach (Anderson, Duvall
\& Jefferies~1990).
Here $\nu_c$ is the central value of $\nu$ in the fitting interval,
$\nu_0$ is the fitted frequency and $w_0$ is the half-width.
The peak power in the mode is given by $\exp(A_0)/w_0^2$. The terms
involving $b_1$ and $b_2$ define the background power.
We generally fit each peak in the
spectra separately by fitting a region extending roughly
half way to the adjoining peaks. In order to check the sensitivity of
fits to the presence of neighboring
peaks we also fit two or three neighboring modes simultaneously.
We find that these
independent fits give results which agree with each other to
within the estimated errors.
We  examine each of the fits visually to reject all those
cases where the program fitted  stray noise spikes in the 
power spectra. Apart from this, we also fit the 3d spectra
directly as  described by Basu et al.~(1999).
We find that  these frequencies also
agree with those determined from averaged 2d spectra.
For the full-disk spectra we have also fitted the sum of the 24
spectra centered at the equator (and $\pm15^\circ$ latitude)
and these frequencies also agree with those
calculated from the sum over all 72 spectra.
Thus it is clear that the fitted frequencies are fairly robust.

\section{Results}

Fig.~1 shows the $\ell$-$\nu$ diagram for the fitted modes. This includes
the results obtained from all three spectra. It is clear that the
frequencies computed from the three independent spectra agree with
each other as all points appear to fall in the same series of ridges.
These frequencies are also in
agreement with those determined by Rhodes et al.~(1998) from MDI spectra
for $\ell<1000$ as well as with those determined by
Fernandes et al.~(1992).
This figure also shows the frequencies computed for
two  solar models, one  without a chromosphere (henceforth called
Model 1) and one with a  chromosphere (Model 2). 

Both Model 1 and Model 2 are static solar models
computed using the hydrogen abundance profile inferred from helioseismic
data (Antia \& Chitre 1998) and heavy element abundance profile from
model 5 of Richard et al.~(1996). These models use OPAL opacity tables
(Iglesias \& Rogers 1996), OPAL Equation of state (Rogers, Swenson
\& Iglesias 1996) and nuclear reaction rates as adopted by
Bahcall \& Pinsonneault~(1995). We use the formulation of Canuto \&
Mazzitelli~(1991) to calculate the convective flux. 
Thus both models are identical in the interior, but Model~1 extends only
up to the temperature minimum, while Model 2 includes a chromosphere.
For the
chromospheric region in Model 2 we use Model~C of Vernazza, Avrett \&
Loeser~(1981). 

The frequencies for these solar models are computed
using the usual adiabatic equation, neglecting all dissipative effects.
For Model 1, the outer boundary conditions for calculating the 
frequencies are applied at the temperature minimum, for Model 2,
they are applied in the chromosphere corona transition layer
at  a height of 2300 km above the solar surface.
The frequencies of Model 1 (no chromosphere)  show the normal f- and
p-mode ridges. For Model 2, which has a chromosphere,  we get a series of
avoided crossings between these p-mode ridges, corresponding to each
chromospheric mode (cf., Ulrich \& Rhodes 1977).
In addition we also get a series of chromospheric
g-modes for this model.

We show the width of the fitted modes in Fig.~2. The figure also shows
the locus of points
where the damping time ($1/2\pi w_0$) equals the period of
oscillatory mode. The width for f- and $p_1$-modes with $\ell<1800$
agrees reasonably with the values obtained by Duvall et al.~(1998).
It may be noted that the
half-width of p-modes appears to reach a constant value of about
0.2 mHz at high frequencies ($\nu>6$ mHz).
The corresponding life-time
for these modes is comparable to the wave travel time 
from surface to lower turning point and back (cf., Duvall et al.~1991).
Most of the scatter seen in Fig.~2 at high frequencies is due to
statistical errors in estimating these widths. From the figure it
appears that at low frequencies the width increases with decreasing
frequencies. We believe that this is artificial and due to the fact
that the  width is actually less than the resolution limit of our spectra.
We have tried fits keeping the width fixed for these modes to check
that the fitted frequencies are not affected by this behavior.

Fig.~3 shows the power spectra for $\ell=1295$ obtained using the full-disk
Dopplergrams. Apart from the normal f- and p-modes it shows two more
peaks at frequencies of about 1.25 and 2.35 mHz and some other peaks 
between the p-mode peaks. These peaks are not seen in the spectra obtained
using high-resolution Dopplergrams and they are believed
(Duvall, Hill \& Scherrer; private communications), to be due
an artifact arising from the fact that tracking the full-disk image
with solar rotation rate leads to a shift by one pixel in approximately
800 s, giving a frequency of 1.25 mHz. This frequency then produces
side lobes for each of the f- and p-modes giving rise to other peaks
with frequencies differing by $\pm 1.25$ mHz from the real peaks.

\subsection{The f-mode}

As can be seen from Fig.~1, we are able to detect the f-mode ridge
up to $\ell\approx4000$.
The f-modes are surface gravity modes and their frequencies are
asymptotically expected to satisfy the simple dispersion relation
\be
\omega^2\approx gk= {GM_\odot\sqrt{\ell(\ell+1)}\over
R_\odot^3}, \label{asym}
\ee
where $g$ is the acceleration due to gravity at
the surface, $k$ is the horizontal wave number, $G$ is the gravitational
constant and $\ell$ the degree of the  mode.
The f-mode frequencies have previously been determined  using different data sets
(e.g., Bachmann et al.~1995; Duvall et al.~1998)
and some of these results are compared with our measurement in Fig.~4.
This figure shows $\omega^2/gk$ evaluated at the surface ($r=R_\odot$). 
Theoretically, this ratio is expected to be around unity for f-modes.
We find that the frequencies obtained from the full-disk and the
high-resolution observations are close to each other,
though there is a small systematic difference between the two.
This difference could be due to error in estimating the spatial scale
in high-resolution Dopplergrams.
There is also some systematic difference between frequencies obtained
from different observations which should
give an estimate of systematic errors that can be expected
in these frequencies.
All sets of observed frequencies show a similar dip in the curve.
For comparison, we also show this ratio for Model 2.

The dip in the curve for the model is due to the fact that for large
$\ell$ the f-modes are effectively trapped in the atmosphere where
$gk\sim1/r^3$ is smaller than its surface value.
This happens because f-modes are
characterized by eigenfunctions where the radial velocity $v_r$, increases
outwards exponentially  as $e^{kz}$, where 
$z$ is the height above the solar surface (Gough 1993).
For small $k$, the velocity scale height is much larger than the density
scale height in the solar atmosphere and hence the kinetic energy density
$\rho v_r^2$ decreases with height. Thus these modes are localized
below the solar surface.
But for $\ell\ga2000$
the kinetic energy density may not decrease with height near the top
of convection zone and the peak in kinetic energy density will
shift outward towards the chromosphere-corona transition region. 
However, the minimum value of density scale height in
chromospheric models is about 100 km, and hence for f-modes with
$\ell\ga3500$ the kinetic energy density may keep increasing with height
and we would not expect such modes to be observed.
In Model 2 we have imposed an artificial boundary at a height
of about 2300 km ($0.0033R_\odot$), and hence 
$\omega^2$  saturates to $gk$ corresponding to this height.
The Sun, however, does not have a well defined boundary and we do not
expect this behavior at large $\ell$.

The observed $\omega^2/gk$
shows a sharp dip at intermediate values of $\ell=1000-2000$ 
which Duvall et al.~(1998) have interpreted as being due to turbulent
velocity fields in the solar convection zone.
We find that there is a distinct rise in the ratio $\omega^2/gk$
at higher $\ell$ ($\ga2500$) which is not properly understood.
This may be due to the fact that these modes are located in
the chromosphere rather than subsurface
layers, where the relevant velocity fields will be different.

It can be seen from Fig.~2 that the width for f-modes increases much
more rapidly with $\nu$ as compared to the p-modes. Further, the width
becomes very large by
$\ell\approx3000$ ($\nu\approx 5.35$ mHz)
and the corresponding damping time ($1/2\pi w_0$)
is comparable to the time-period of the oscillation. 
Hence beyond this point the mode can
not be considered to exist as a normal standing wave.
This is expected since beyond this value of $\ell$, the kinetic energy
density for f-mode will not decrease with height. The sharp increase
in the width of these modes is probably a reflection of that.
At even higher degrees, it appears that the width becomes roughly constant
with increasing frequency and we believe that the mode is no longer an
f-mode.  From Fig.~1 it appears that at very high $\ell$
the frequencies of the f-modes tend towards those of the
chromospheric modes  in Model 2. It may be noted that because of very
large half-width ($\approx 1$ mHz) and low power it is not possible
to separate the f-mode ridge from  the lowest chromospheric mode ridge
in this region. It is quite possible that for $\ell\ga3000$ the f-mode
power reduces sharply and hence the corresponding ridge in the
observed spectra is dominated by
that of the chromospheric mode. This probably explains the sharp increase
in $\omega^2/gk$ seen in Fig.~4 and also the fact that the width appears
to become roughly constant for $\ell\ga3000$.

\subsection{Influence of the chromosphere on p-modes}

It appears from Fig.~1 that the frequencies of Model 1
are in remarkable agreement with the observed p-mode ridges.
It is possible that this agreement is merely a coincidence where
the effect of dissipation has compensated the expected shift from
inclusion of chromosphere. To test this we can try to estimate the
phase shift associated with reflection of the wave at upper turning
point for these modes by fitting the frequencies
to the asymptotic form 
\be
{\pi(n+\alpha(\omega))\over\omega}=F\left({\omega\over L}\right),
\ee
where $L=\ell+1/2$ and $\alpha(\omega)$ is believed to
include the contribution arising from phase shift at outer turning
point (cf., Duvall et al.~1991).
We have estimated $\alpha(\omega)$ by fitting this form in the
frequency interval of 2--8.5 mHz and the results are shown in Fig.~5.
It is clear that there is a significant difference between
the observed and model frequencies at intermediate frequencies.
For observed frequencies $\alpha$ changes rather steeply at
intermediate frequencies, which is where the upper turning point
for modes shifts from photosphere to chromosphere corona transition
region. This is not the case for frequencies of Model~1.
Further, for observed frequencies $\alpha$ is almost constant
for $\nu\ga5$ mHz. This can be considered to be an estimate of the acoustic
cutoff frequency in the photosphere. Above that frequency all modes
propagate in the chromosphere, as a result these modes
suffer roughly the same phase shift
at the upper turning point which should be located in the chromosphere
corona transition layer.

Fig.~6 shows the frequency separation between adjacent modes of the
same degree. For observed frequencies $(\nu_{n+1,\ell}-\nu_{n,\ell})$
shows a bump around $\nu=5-6$ mHz.
This could be a signature of avoided crossings between these modes,
which have not been explicitly seen because of low power in the
chromospheric modes. The amplitude of the bump
decreases with increasing $n$, which is expected since modes of the
same frequency but larger $n$ will penetrate deeper below the surface
making the relative effect of chromosphere smaller.
The bump also shifts towards lower frequency as $n$ increases,
which is to be expected as the frequency of the second chromospheric
mode in Model~2 does decrease slowly as $\ell$ reduces (cf., Fig.~1).
The frequency separation for Model 1 does not show any such bump.
The first chromospheric mode around a frequency of 4 mHz
has a  relatively sharp avoided crossing, as seen for the
theoretical solar model and is
not likely to cause any significant bump in the frequency separation.
At higher frequencies, the chromospheric cavity merges with the
internal cavity and there would be no separate chromospheric modes.
Hence we do not expect any more bumps in the frequency separation.

\subsection{Effect of asymmetry of the peak profiles}

From Fig.~3 it is clear that the profile for f-mode peak is significantly
asymmetric with higher power on the lower side of the peak.
Hence, fit to a symmetric Lorentzian profile given by Eq.~(1)
is likely to result in significant shift in the fitted frequencies.
In order to get better fit we have also attempted fits using
asymmetric profiles of the form (cf., Nigam \& Kosovichev 1998)
\be
P(\ell,\nu)={e^{A_0}[(1+Bx)^2+B^2]\over1+x^2}+
e^{b_1}[1+b_2(1-\nu/\nu_c)],
\ee
where $x=(\nu-\nu_0)/w_0$ and $B$ is a parameter that controls the
asymmetry. This parameter is positive for positive asymmetry, i.e., more
power on the higher frequency side of the peak, and
negative for negative asymmetry. The parameter $B$ is also fitted along
with other parameters defining the line profile.
For low order modes ($n=0,1,2$) this
parameter is found to be significant, while at higher radial order it
is difficult to distinguish between fits with symmetric or asymmetric
profiles and the value of $B$ is generally within the corresponding
error estimates. 

Fig.~7 shows the fits for $\ell=1195$ f-mode using both symmetric and
asymmetric profiles.  It is clear from this figure that the symmetric
profile does not fit the observed power spectrum properly and tends
to underestimate the frequency. The asymmetric profile leads to a nearly
perfect fit to the observed spectrum and we would expect the corresponding
frequency to be more reliable. Since $B$ is negative for all f-modes
the frequency shifts to higher value when asymmetric profile is used
for fitting.

Fig.~8
shows $\omega^2/gk$ for f-mode frequencies obtained using the symmetric
and asymmetric profiles. It is clear that use of asymmetric profiles
results in frequencies which are significantly higher and the shift
in frequencies due to use of asymmetric profiles is much larger than the
error estimate in individual fits. Further, the difference between the
observed and model frequencies for f-modes is somewhat reduced
when asymmetric profiles are used. However, it is clear from the figure that
even after accounting for asymmetry in profiles, the observed frequencies
around $\ell=2000$ are significantly lower than the frequencies computed
for a solar model.

Fig.~9 shows the asymmetry parameter $B$ as obtained from the fits using
asymmetric profiles for $n=0,1,2$. It is clear that the value of $B$
is significantly different from zero, which is expected because of clear
asymmetry seen in power spectra (Figs.~3, 7). For $p_2$-mode at low
frequency it is not possible to distinguish between fits to symmetric
and asymmetric profiles and the resulting value of $B$ is consistent with
zero within the estimated errors. The same is true for higher radial order
$n$ and hence these modes are not fitted using asymmetric profiles as the
fits with symmetric profiles are found to be more stable.
The resulting scaled frequency shift between the asymmetric and symmetric
profiles is shown in Fig.~8. The frequency differences are scaled by
$Q_{n\ell}$, the ratio of the inertia of the mode to that of a radial
mode with the same frequency (\jcd\ \& Berthomieu 1991).
It appears that to first approximation the
scaled frequency shift is a function of frequency alone. The net shift
in f-mode frequency is much larger than those in $p_1$-mode, but because
of difference in $Q_{n\ell}$ the scaled frequency differences are of same
order. For f-modes the frequency shift increases with $\ell$ from
less than 1 $\mu$Hz at $\ell\approx250$ to 
over 100 $\mu$Hz at $\ell\approx3000$.
Further, for p-modes the frequency shifts are comparable to
the corresponding error estimates. Nevertheless, the scaled frequency
shifts are not exactly a function of frequency alone and hence the
correction in frequencies due to asymmetric profile can affect the
results of helioseismic inversions to some extent.

\section{Conclusions}

Using the ring diagram technique applied to MDI data we have
determined the frequencies of solar oscillations covering
a wide range of degree and frequency. We have improved  the quality 
of the  power spectra by taking the  sum over several spectra, thus
effectively increasing the time duration of observations to about 26 days.
The line widths for the high frequency
($\nu>6$ mHz) p-modes appear to be roughly constant, which is to be
expected as all these modes have significant amplitude in the
chromosphere, where radiative damping is dominant.

The f-mode frequencies at high wavenumbers deviate
significantly from those calculated for a solar model.
The  ratio $\omega^2/gk$ for the observed frequencies 
shows a sharp dip around $\ell=2500$, after which this ratio increases.
It appears that the
f-mode does not have significant power beyond $\ell=3000$, which is
consistent with the fact that the minimum scale height in the solar
atmosphere is around 100 km. The damping time estimated from the width
of f-mode becomes shorter than the time-period around $\ell=3000$
and the apparent continuation of the ridge at higher degree is probably
the lowest chromospheric p-mode.

The f-mode peaks in power spectra are distinctly asymmetric and use of
symmetric profiles in fitting the power spectrum results in significant
shift in the estimated frequencies. The estimated frequencies increase
by up to 100$\mu$Hz, when asymmetric profiles are used in fitting.
The frequency shift due to asymmetric profiles reduces at low degree
but it may still be significant for $100\le \ell\le250$ which has been
studied using global power spectra (Schou et al.~1997;
Antia 1998). This may also affect the seismic estimate of solar
radius.

Although we do not find any direct evidence for the avoided crossings
expected due to chromospheric modes, the frequency difference
$\nu_{n+1,\ell}-\nu_{n,\ell}$ has a distinct bump around a frequency
of 5--6 mHz, which may be due to avoided crossings. This is the same
frequency range where Harvey et al.~(1993) and Tripathy \& Hill~(1995)
have observed a broad background feature in the normal p-mode spectra.

\acknowledgments

This work  utilizes data from the Solar Oscillations
Investigation / Michelson Doppler Imager (SOI/MDI) on the Solar
and Heliospheric Observatory (SOHO).  SOHO is a project of
international cooperation between ESA and NASA.
The authors would like to thank the SOI Science Support Center
and the SOI Ring Diagrams Team for assistance in data
processing. The data-processing modules used were
developed by Luiz A. Discher de Sa and Rick Bogart, with
contributions from Irene Gonz\'alez Hern\'andez and Peter Giles.
We thank  F. Hill, P. Scherrer and T. Duvall for pointing out
the 1.25 mHz artifact in power spectra, which were incorrectly interpreted
as chromospheric modes in earlier version of the manuscript.

\vfill\eject

\begin{figure}
\plotone{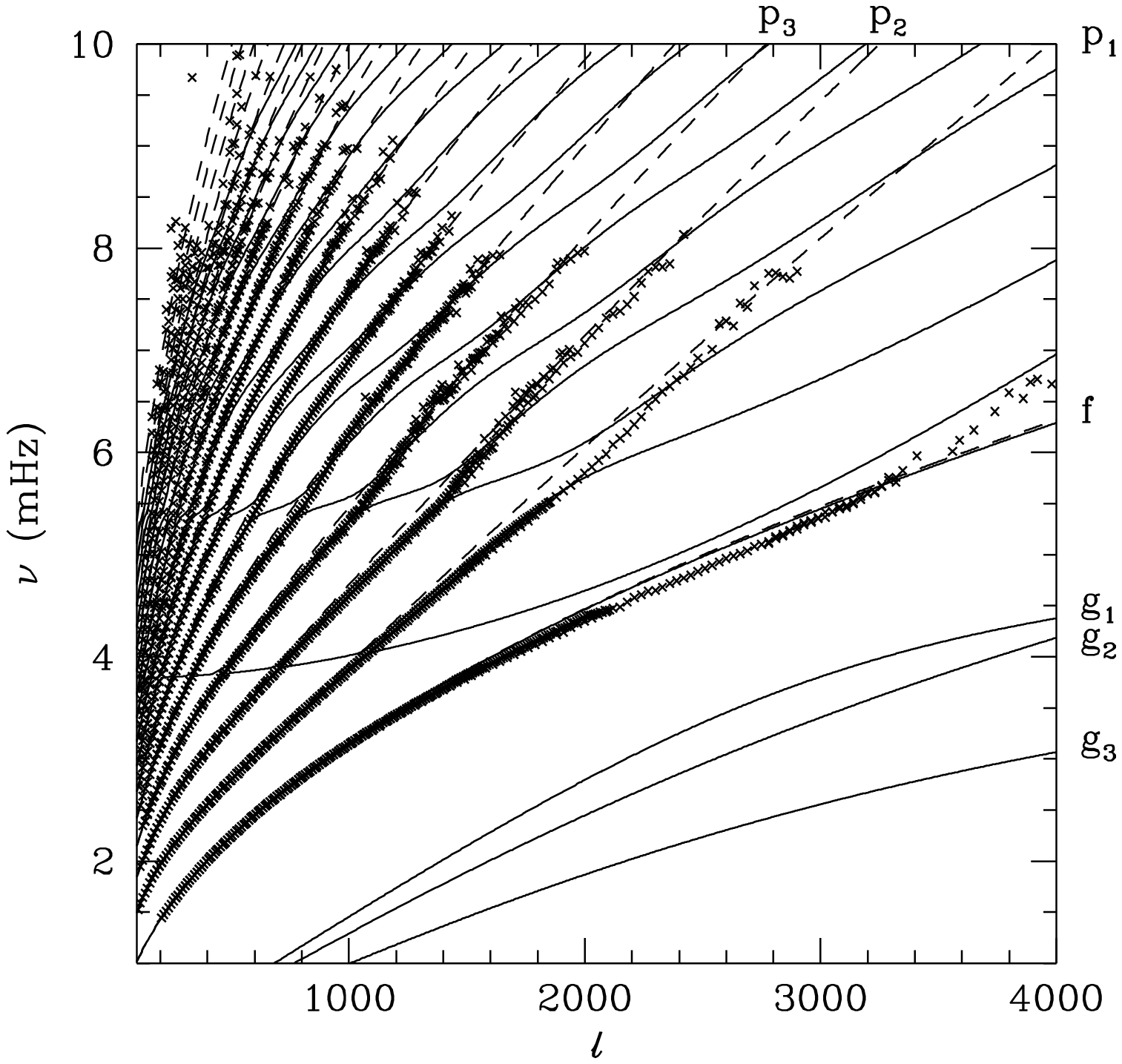}
\figcaption{The $\ell$--$\nu$ diagram of the modes obtained
by our analysis. The crosses display the fitted 
 modes obtained from the full-disk
and high-resolution Dopplergrams, including the full-disk observations
at higher cadence of 30s.
The continuous lines show the modes
for Model 2 (i.e., one which has a chromosphere). The dashed lines are the
ridges for Model 1, which is  truncated at the
temperature minimum. The labels near the right and top axis identify
some of these ridges.
}
\end{figure}

\begin{figure}
\plotone{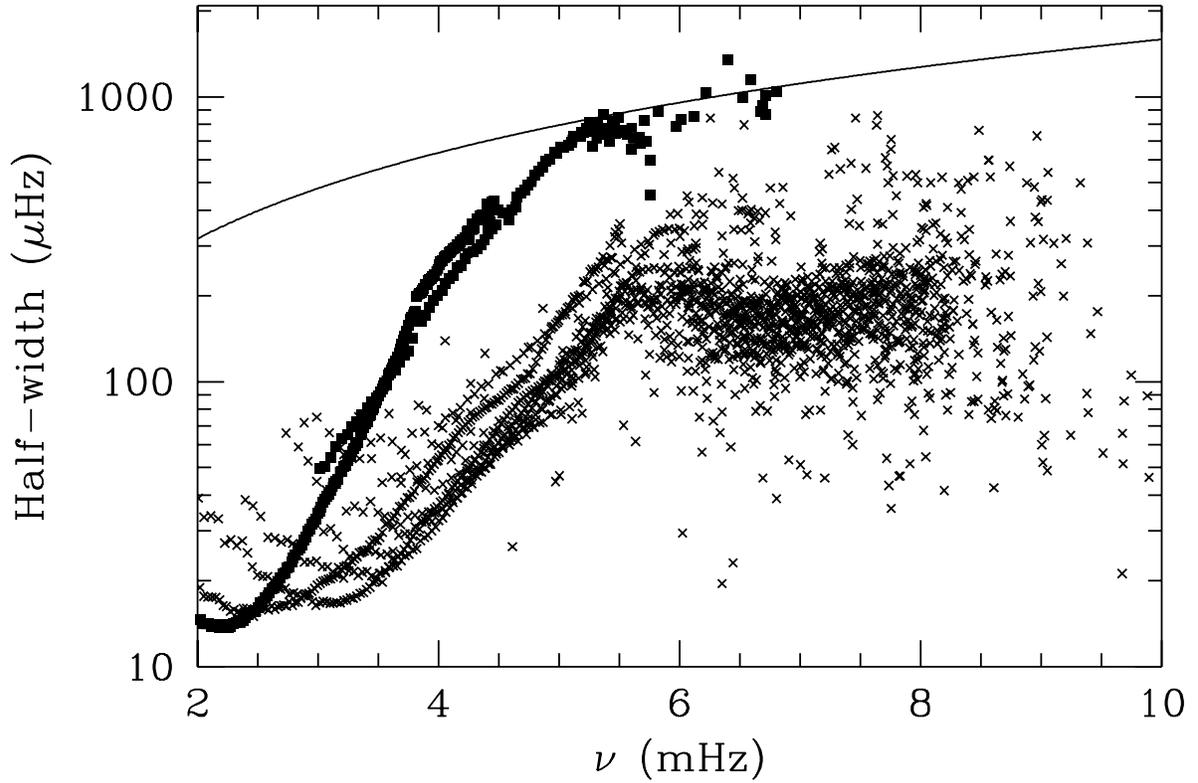}
\figcaption{The half-width of the fitted modes.
The squares show the f-modes while the crosses are for the p-modes.
The continuous line
marks the width for which the damping time is equal to the time-period
of oscillation.
}
\end{figure}

\begin{figure}
\plotone{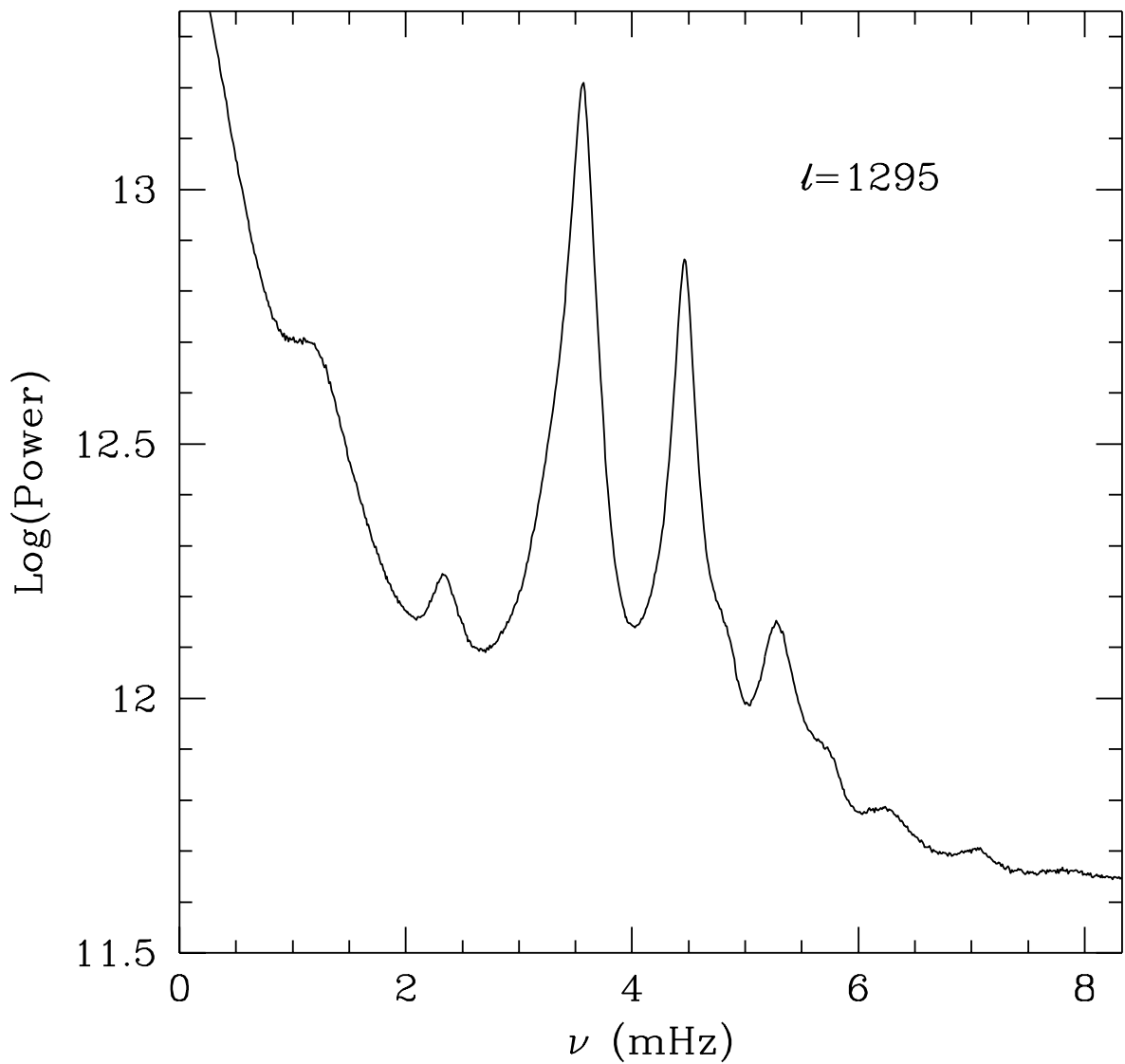}
\figcaption{ The power spectrum for $\ell=1295$ obtained
from full-disk Dopplergrams.
The vertical axis is in arbitrary units. The peaks at around
1.25 and 2.35 mHz are artifacts introduced by tracking of images
with solar rotation rate.
}
\end{figure}

\begin{figure}
\plotone{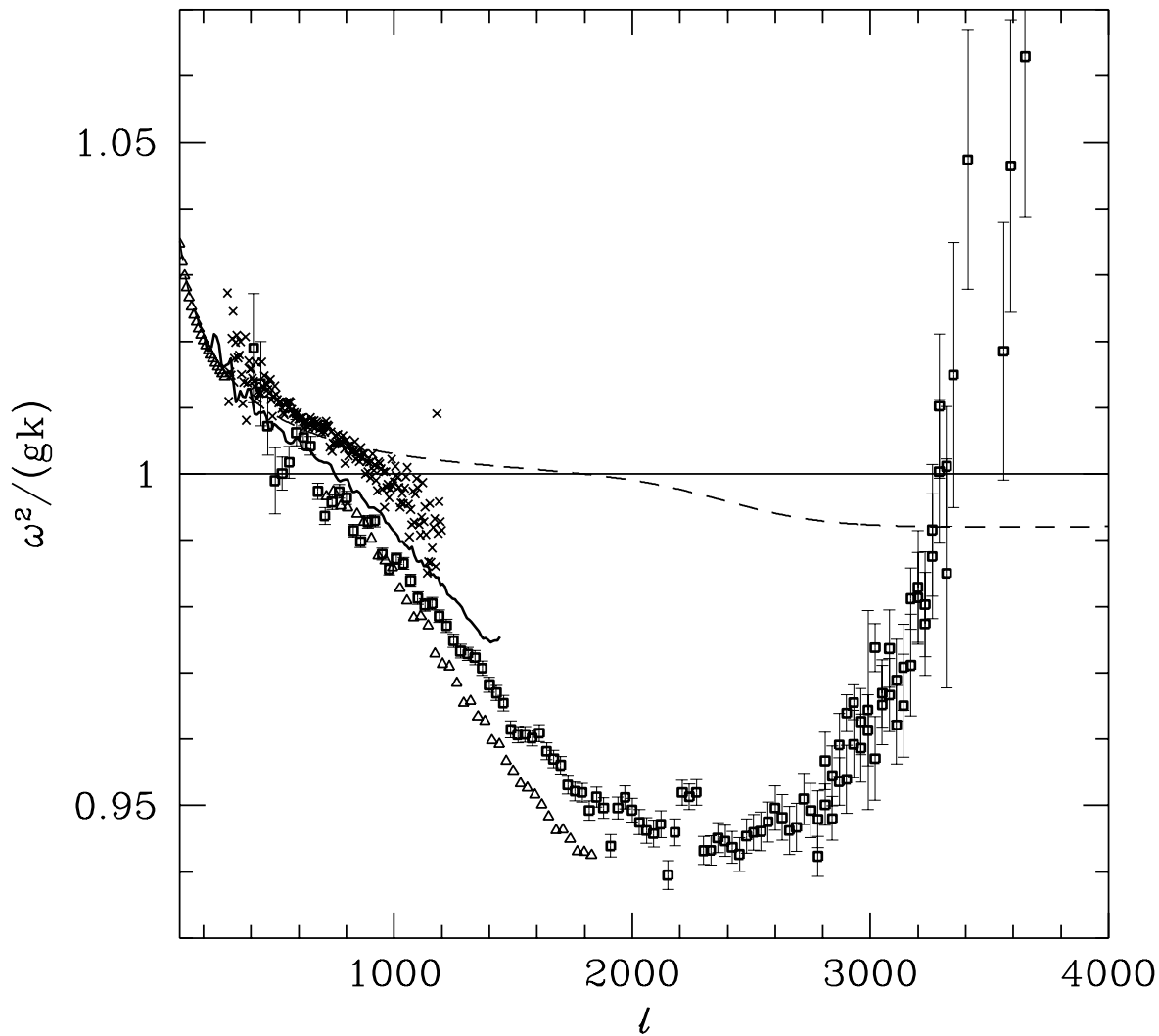}
\figcaption{The ratio $\omega^2/gk$ for various observed
frequencies of f-modes are displayed.
The crosses are data from Bachmann et al.~(1995).
the triangles are from MDI data
(Rhodes et al.~1997; Duvall et al.~1998). 
The continuous line shows the results obtained by us using the full-disk
Dopplergrams while the squares with error bars represent those from the
high-resolution Dopplergrams. The dashed line shows the result for 
Model 2.}
\end{figure}

\begin{figure}
\plotone{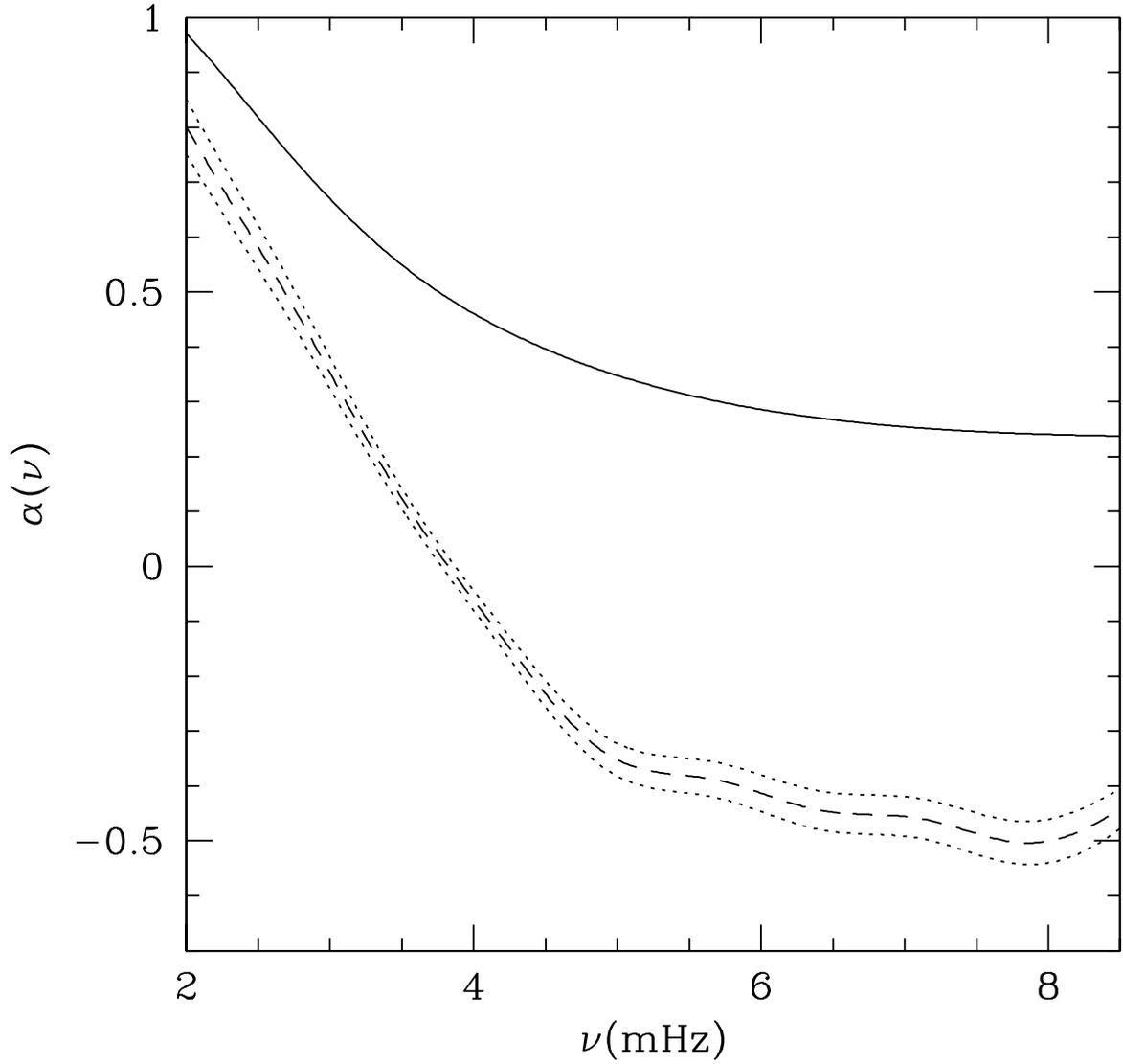}
\figcaption{The phase $\alpha(\nu)$ defined by Eq.~(3) as a
function of frequency. The continuous line is  for Model 1, 
which is  a solar model with upper boundary at the temperature minimum.
The dashed line is for the observed frequencies. The dotted lines show the
error estimates.}
\end{figure}

\begin{figure}
\plotone{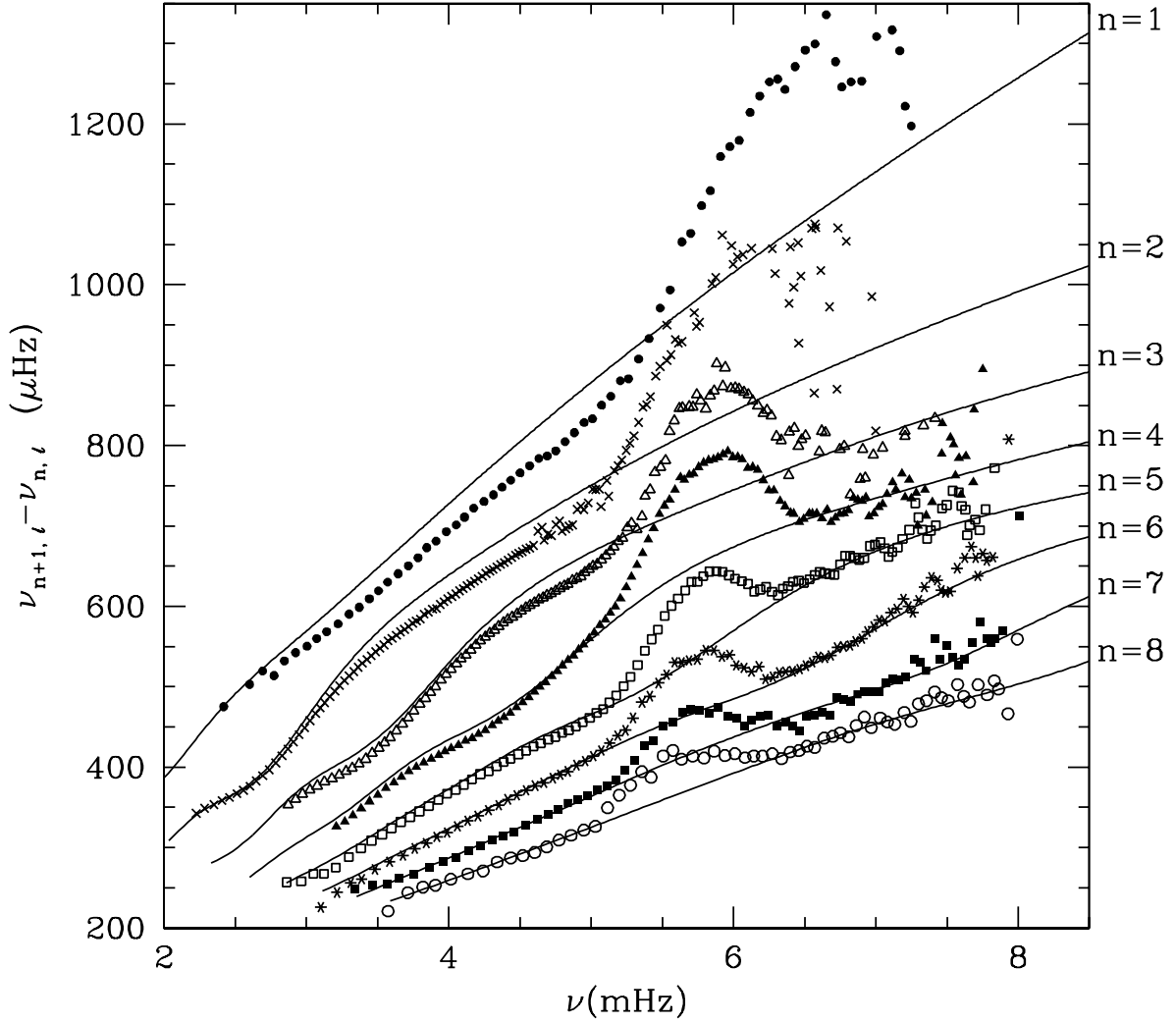}
\figcaption{ The frequency separation between adjacent  modes of the
same degree, $\nu_{n+1, \ell}-\nu_{n, \ell}$, plotted as a function of
the mean frequency of the modes. The lines are the frequency separations
for  Model~1 for different values of $n$
and the points are for the observed frequencies.
The filled circles are the  observed frequency separations for $n=1$,
crosses for $n=2$, empty triangles for $n=3$,
filled triangles for $n=4$, empty squares for $n=5$,  asterisks
for $n=6$, filled squares for $n=7$ and empty circles for $n=8$. }
\end{figure}

\begin{figure}
\plotone{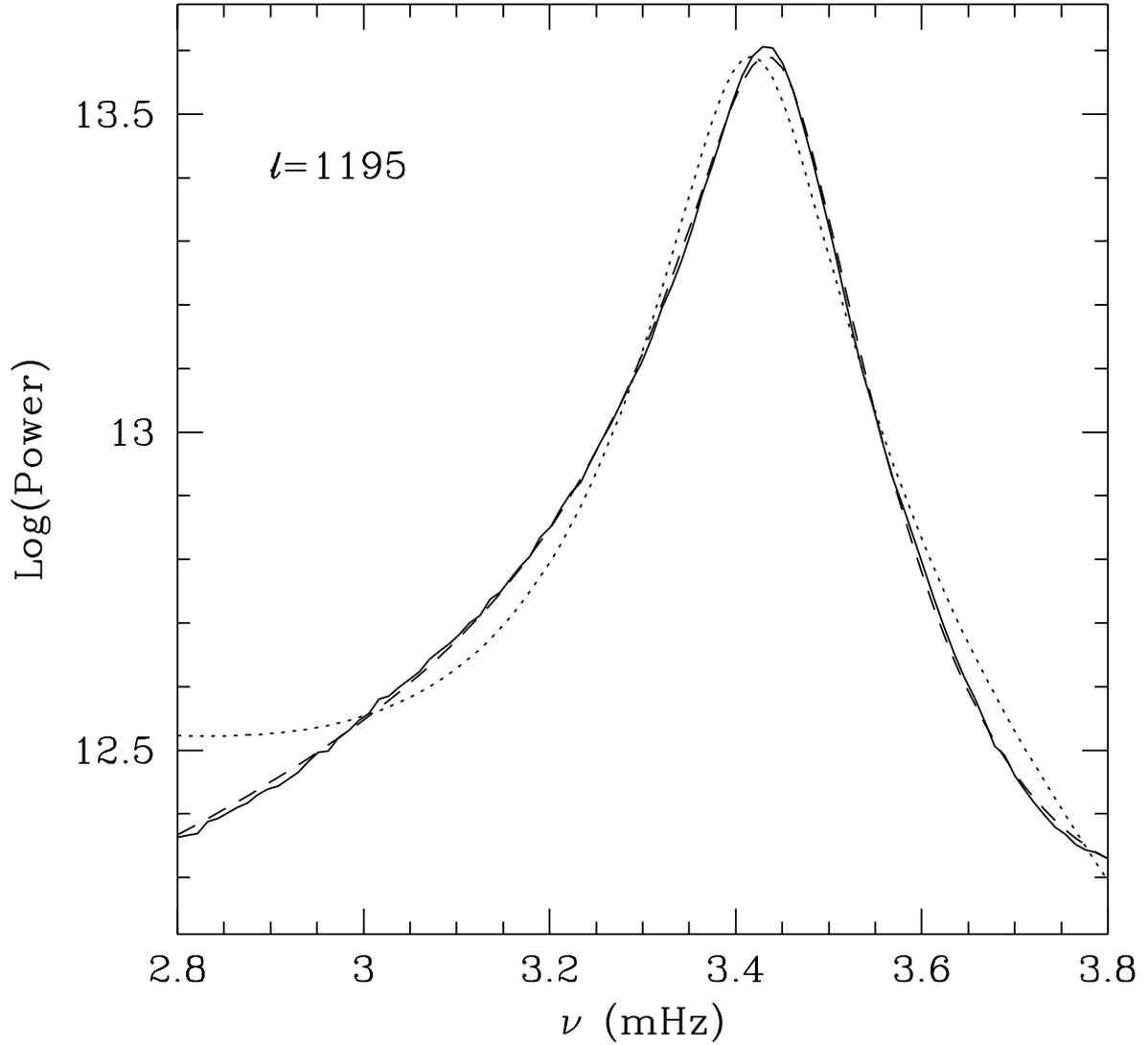}
\figcaption{The fit to f-mode peak in the power spectrum obtained from full-disk
Dopplergrams for $\ell=1195$. The continuous line shows the observed
power spectra, while the dotted line shows the fit using symmetric
Lorentzian profile and the dashed line shows the fit using asymmetric
profile given by Eq.~(4).}
\end{figure}

\begin{figure}
\plotone{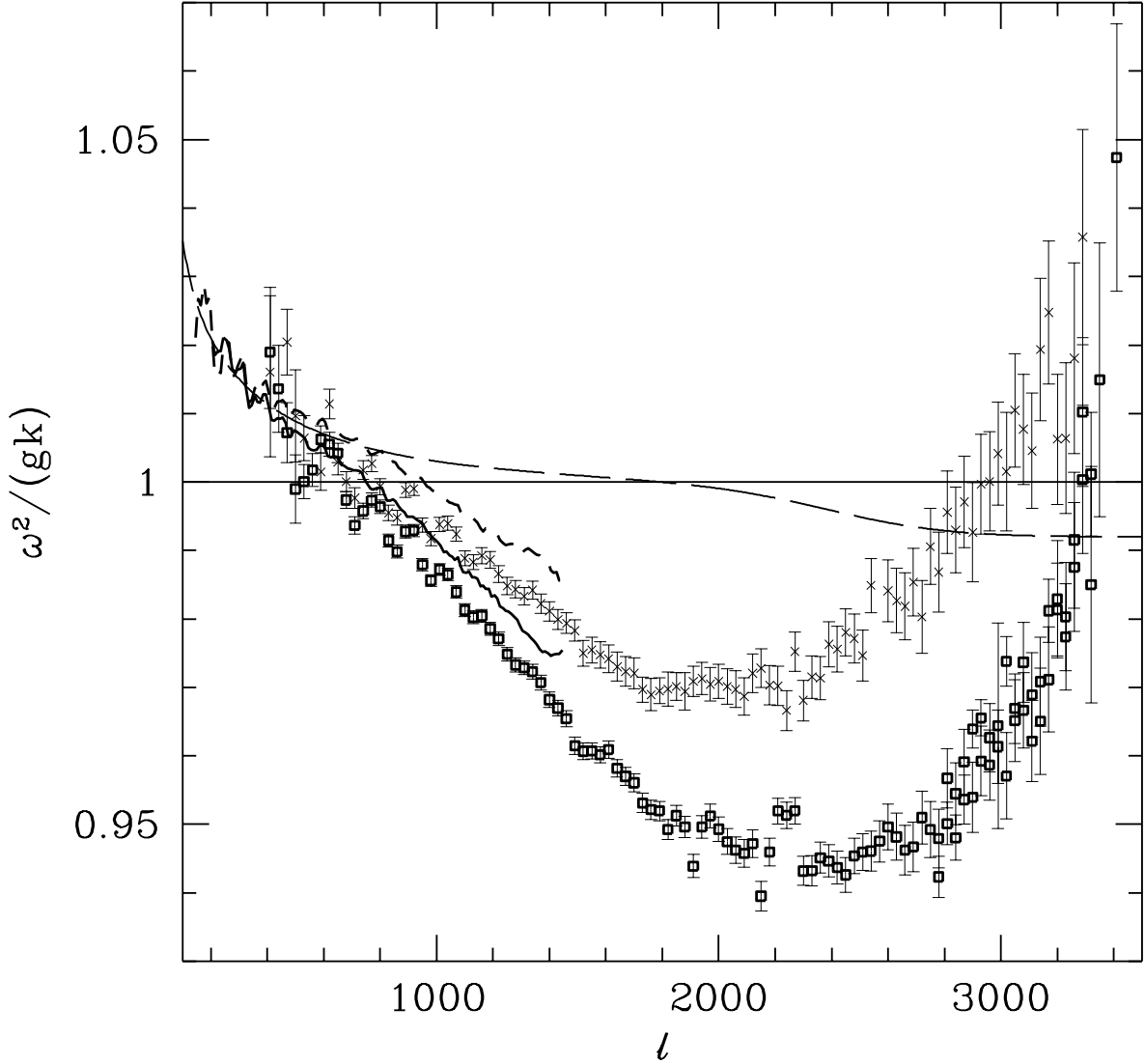}
\figcaption{The ratio $\omega^2/gk$ for various fitted
frequencies of f-modes are displayed. The results obtained using the
full-disk Dopplergrams are shown by the continuous and short-dashed lines
for fits using symmetric and asymmetric profiles respectively.
Squares and crosses with error bars respectively, show the results
obtained using symmetric and asymmetric profiles for the spectra obtained
using high-resolution Dopplergrams.
The long-dashed line shows the result for Model 2.}
\end{figure}

\begin{figure}
\plotone{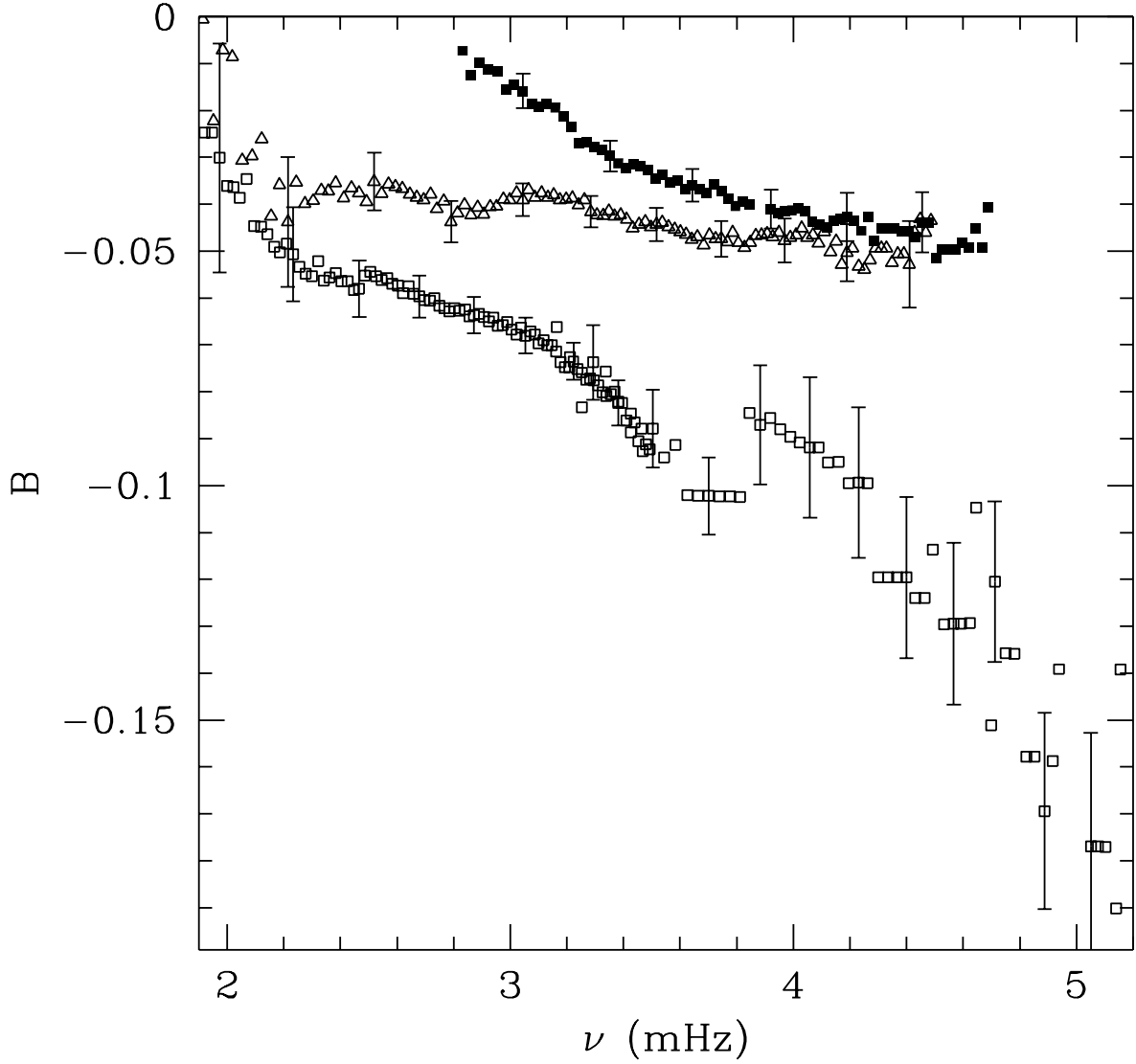}
\figcaption{The asymmetry parameter $B$ for $n=0$ (open squares),
$n=1$ (triangles) and $n=2$ (filled squares). 
For clarity error-bars are shown for a few modes only.}
\end{figure}

\begin{figure}
\plotone{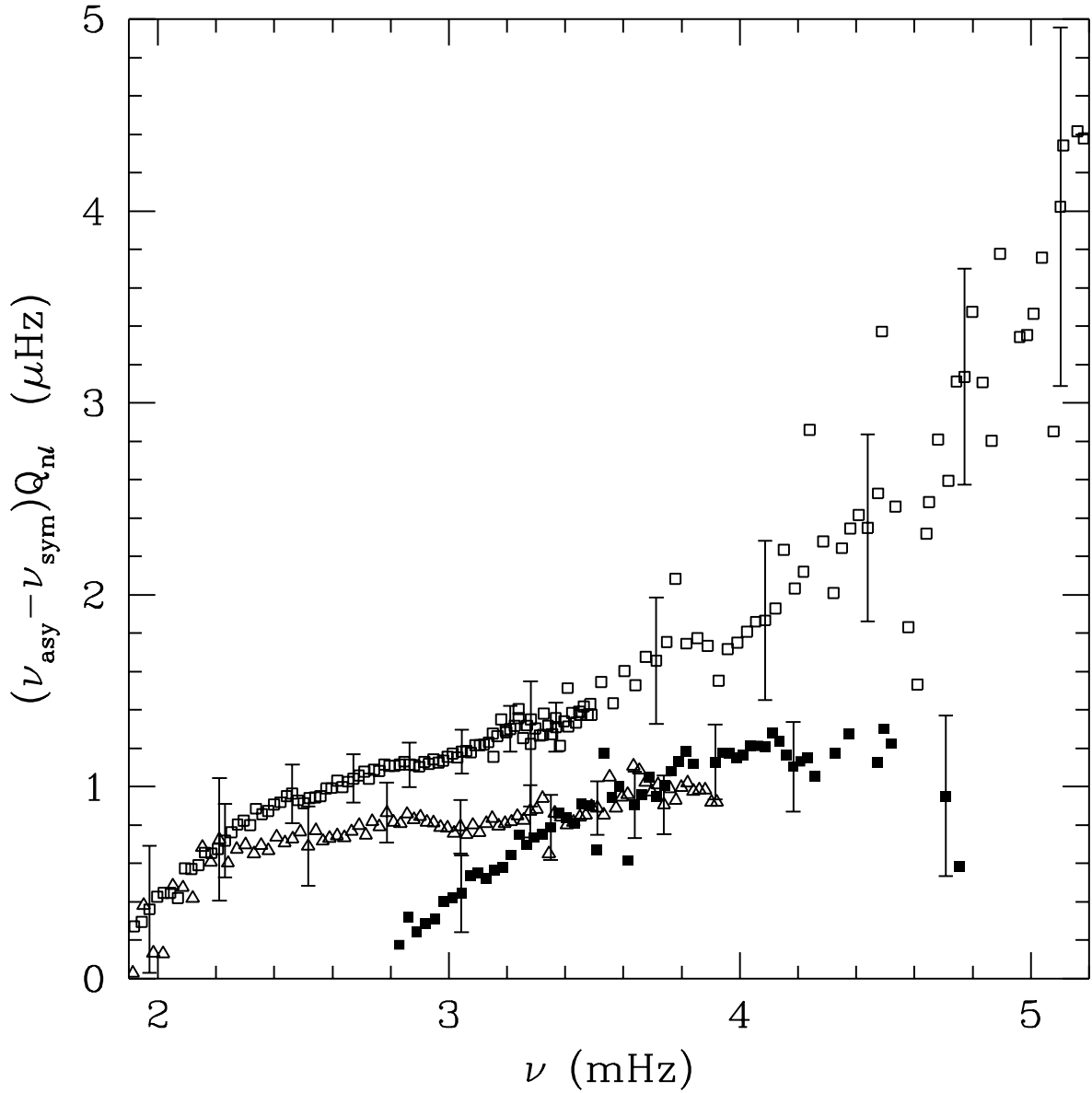}
\figcaption{The scaled frequency shifts for $n=0$ (open squares),
$n=1$ (triangles) and $n=2$ (filled squares).
For clarity error-bars are shown for a few modes only.}
\end{figure}

\end{document}